\documentclass[prl,aps,twocolumn,groupedaddress,showpacs,floatfix,preprintnumbers]{revtex4}
\usepackage{amsmath,amssymb,multirow,bm}

\usepackage{bm}
\usepackage{epsfig}
\usepackage{amsbsy}
\usepackage{dcolumn}
\usepackage{stackrel}

\newcommand{\be}{\begin{equation}}
\newcommand{\ee}{\end{equation}}
\newcommand{\bea}{\begin{eqnarray}}
\newcommand{\eea}{\end{eqnarray}}

\newcommand{\avg}[1]{\left \langle \psi \left | #1 \right | \psi \right \rangle }
\newcommand{\avgo}[1]{\left \langle \psi_0 \left | #1 \right | \psi_0 \right \rangle }
\newcommand{\pictsize}{1.0}
\newcommand{\ssec}[1]{\emph{#1}.---}

\begin{document}
\preprint{NT@UW-14-07}

\title{Auxiliary-Field Quantum Monte Carlo Simulations of Neutron Matter in \\ Chiral Effective Field Theory}

\author{G.\ Wlaz{\l}owski$^{1,2}$, J. W.\ Holt$^2$, S.\ Moroz$^2$, A.\ Bulgac$^2$,   K. J.\ Roche$^{2,3}$}
\affiliation{$^1$Faculty of Physics, Warsaw University of Technology, Ulica Koszykowa 75,  00-662 Warsaw, Poland}
\affiliation{$^2$Department of Physics, University of Washington, Seattle, WA 98195, USA}
\affiliation{$^3$Pacific Northwest National Laboratory, Richland, WA 99352, USA}

\begin{abstract}
We present variational Monte Carlo calculations of the neutron matter equation of state using 
chiral nuclear forces. 
The ground-state wavefunction of neutron matter, containing non-perturbative 
many-body correlations, is obtained from auxiliary-field quantum Monte Carlo simulations of up 
to about 340 neutrons interacting on a $10^3$ discretized lattice. The evolution Hamiltonian is chosen 
to be attractive and spin-independent in order to avoid the fermion sign problem and is 
constructed to best reproduce broad features of the chiral nuclear force. This is facilitated 
by choosing a lattice spacing of 1.5 fm, corresponding to a momentum-space cutoff of $\Lambda = 
414$ MeV/c, a resolution scale at which strongly repulsive features of nuclear two-body 
forces are suppressed. Differences between the evolution potential 
and the full chiral nuclear interaction (Entem and Machleidt $\Lambda = 414$ MeV) are then treated 
perturbatively. Our results for the equation 
of state are compared to previous quantum Monte Carlo simulations 
which employed chiral two-body forces at next-to-next-to-leading order (N2LO). In addition
we include the effects of three-body forces at N2LO, which provide important repulsion at 
densities higher than 0.02 fm$^{-3}$, as well as two-body forces at N3LO.
\end{abstract}

\date{\today}


\maketitle
% 21.65.Cd	Asymmetric matter, neutron matter
% 21.30.-x	Nuclear forces (see also 13.75.Cs Nucleon-nucleon interactions)
% 21.60.De	Ab initio methods
% 21.60.Ka	Monte Carlo models

% {\small  Monte Carlo, Keywords: Effective field theory at finite density, Chiral three-nucleon force.}

%%%%%%%%%%%%%%%%%%%%%%%%%%%%%%%%%%%%%%%%
%%%%%%%%%%%%%%%%%%%%%%%%%%%%%%%%%%%%%%%%
% Introduction
%%%%%%%%%%%%%%%%%%%%%%%%%%%%%%%%%%%%%%%%
%%%%%%%%%%%%%%%%%%%%%%%%%%%%%%%%%%%%%%%%
\ssec{Introduction}
Understanding the static and dynamic properties of neutron matter will be key to 
addressing fundamental questions at the interface of nuclear physics and astrophysics. 
The structure and evolution of neutron stars, the identification of viable sites for $r$-process 
nucleosynthesis, and the interpretation of observed gravitational waveforms from compact 
binary mergers depend on neutron matter response functions and the equation of state. 
The nuclear densities relevant in these phenomena range from dilute neutron matter 
($\rho \simeq 0.0005$\,fm$^{-3}$), governed largely by the universal properties of unitary 
Fermi systems, to several times nuclear saturation density ($\rho_0 \simeq 
0.16$\,fm$^{-3}$) found in the core of neutron stars. Due to the large neutron-neutron
scattering length, low-density neutron matter is tractable through nonperturbative many-body
methods~\cite{Gandolfi09,Gandolfi2009,Gezerlis10,WlazlowskiNM11,Hagen14},
while in the vicinity of nuclear matter saturation density, the equation of state can
be computed to various degree of accuracy and controlled approximations through a variety of
many-body methods~\cite{Hebeler10,Lovato2012,Baldo2012,Rios2012,Carbone2013,Coraggio13,
HoltJW13a,Ekstrom2013,Baardsen2013,Tews2013,Krueger13}.

Recently, a number of quantum Monte Carlo (QMC) studies \cite{EpelbaumEPJA2009,Gezerlis13,Roggero14} of 
neutron matter have employed microscopic nuclear forces derived within the framework 
of chiral effective field theory (for recent reviews see Refs.\ \cite{Epelbaum09,Machleidt11,
HoltJW13b}). These works have focused on chiral two-body interactions at 
order $(Q/\Lambda_\chi)^3$ (or next-to-next-leading order, N2LO), where $Q$ refers to the
low-energy scale set by the pion mass and nuclear momenta, while $\Lambda_\chi$ is the 
chiral symmetry breaking scale set by, e.g., vector meson masses. 

In the present work we introduce a novel approach to study strongly correlated 
nuclear systems on the lattice employing auxiliary-field quantum Monte Carlo (AFQMC)
simulations free of the fermion sign problem. The method enables the
simulation of a larger number of particles than alternative Monte Carlo
implementations, and it offers an avenue to extend ab-initio many-body
methods into the medium-mass region of the nuclear chart. As an
initial application of the method, we focus on the equation of state
of cold neutron matter at low to intermediate densities computed from
chiral two-body forces at N3LO together with the chiral three-neutron
force at N2LO.

The auxiliary-field quantum Monte Carlo 
simulations are performed free of the fermion sign problem by constructing an 
attractive, spin-independent effective Hamiltonian inspired by one-boson exchange models. 
The extent to which such a potential approximates the qualitative features of realistic chiral 
nuclear forces depends, in part, on the resolution scale at which the nuclear force is constructed. 
Lowering the resolution scale weakens the short-distance repulsion in the 
nucleon-nucleon (NN) interaction \cite{Bogner03,Bogner10,Wendt12}, thereby enhancing the role of 
correlations in the neutron matter ground state that can be generated by such evolution 
Hamiltonians. In contradistinction with Green Function Monte Carlo 
simulations~\cite{Gandolfi09,Gandolfi2009,Gezerlis10} employing Argonne nuclear potentials
where the short-range correlations have an important role, in the present approach the emphasis is on the long-range 
correlations.

Specifically, we consider the chiral nuclear interaction described in
Refs.\ \cite{Coraggio07b,Coraggio13,Coraggio14} with the regulating function
\begin{equation}
f(p,p^\prime) = \exp [ -(p/\Lambda)^{2n}
-({p^\prime}/\Lambda)^{2n} ],  
\end{equation}
where $n=10$ and $\Lambda = 414$\,MeV/c. The value of $n$ is chosen to
be large for consistency with the sharp lattice momentum cutoff. This
high-precision nuclear potential reproduces nucleon-nucleon elastic
scattering phase shifts up to lab energies of 200 MeV with $\chi^2/\text{DOF} = 1.44$~\cite{MachleidtPrivComm}, the properties
of the deuteron, the binding energy and lifetime of $^3$H (with the
inclusion of two-body weak currents), as well as the empirical nuclear
matter saturation point and critical point of the liquid-gas phase
transition \cite{Wellenhofer14}. In comparison the optimized evolution
potential, expressed as a sum of attractive and repulsive Yukawa
interactions, is constrained by NN phase shifts as well as the
perturbative equation of state employing the full chiral nuclear
potential. The interacting ground state is then obtained from this
Hamiltonian by propagating a trial Slater-determinant wavefunction in
imaginary time using standard auxiliary-field quantum Monte Carlo
techniques~\cite{Koonin97,Bulgac08}. The expectation value of the full
chiral Hamiltonian in the evolved ground state on the one hand gives
an upper bound on the equation of state and on the other hand can be
interpreted as the first-order perturbative correction in powers of
the difference between the full chiral interaction and the evolution
potential. The present approach establishes the framework for future
work directed toward accessing nucleon spectral properties, linear
response and various transport properties of dilute neutron matter,
similar to what has been demonstrated in the case of the unitary Fermi
gas~\cite{GW2013,Gabriel2013,Gabriel2012,Piotr2011,Piotr2009}.
Spin response and neutrino scattering and
emissivity~\cite{shen2013,shen2013a} as well as collective modes in
dilute neutron matter \cite{chamel2013} are examples of neutron star
and supernova properties that can be addressed.

%%%%%%%%%%%%%%%%%%%%%%%%%%%%%%%%%%%%%%%%%%%
%%%%%%%%%%%%%%%%%%%%%%%%%%%%%%%%%%%%%%%%%%%
% Auxiliary-field quantum Monte Carlo simulations on the lattice
%%%%%%%%%%%%%%%%%%%%%%%%%%%%%%%%%%%%%%%%%%%
%%%%%%%%%%%%%%%%%%%%%%%%%%%%%%%%%%%%%%%%%%%

\ssec{Auxiliary-field quantum Monte Carlo simulations on the lattice}
%\label{afqmc}
Quantum Monte Carlo approaches rely on the very simple idea of projecting out the ground 
state $\psi$ of a many-body system with Hamiltonian $\hat{H}$ by means 
of imaginary time evolution 
% \be
$
\exp(-\tau \hat{H})\psi_{0}\, \stackrel{\tau \rightarrow \infty}{\longrightarrow} \, \psi,
$
% \ee 
where $\psi_{0}$ is an arbitrary initial state with non-vanishing overlap with the ground state. 
In practical realizations the projection is performed by successive application of the evolution 
operator for small imaginary time steps: $\psi(\tau+\Delta \tau)=\exp(-\Delta \tau \,\hat{H})\psi(\tau)$. 
This short evolution in imaginary time is converted into integral form, and the 
emerging multidimensional integration is performed by means of Monte Carlo techniques.
For fermionic systems one has to introduce a prescription for avoiding the sign problem.
The most popular approaches are
the ``fixed-node'' and ``fixed phase'' approximations~\cite{Carlson99,Ortiz93}, where the first one results in a variational approximation
to the energy.
In this paper we utilize a different strategy to deal with 
the sign problem for a large class of systems, which by construction also results in a 
variational estimate of the energy.

Our aim is to compute the ground state energy of the Hamiltonian
% \be
$\hat{H} = \hat{T} + \hat{V}$,
% \ee
where $\hat{T}$ is kinetic energy operator and
% \be
$\hat{V} = \hat{V}_{2N}+\hat{V}_{3N}+\cdots$
% \ee
is the sum of two- and many-body forces. 
In the following we present calculations including chiral 2N interactions up to order N3LO 
in addition to the 3N interaction at order N2LO:
% \be
$
\hat{V}=\hat{V}_{2N}^{(\mathrm{N3LO})}+\hat{V}_{3N}^{(\mathrm{N2LO})}.
$
% \ee
We work with a low-momentum chiral potential with cutoff parameter $\Lambda=414\,\mathrm{MeV/c}$ 
and a steep regulator function~\cite{Coraggio14}. Since the imaginary time evolution of a wavefunction 
with the full chiral Hamiltonian results in a severe sign problem, we rewrite the Hamiltonian as
\be
\hat{H} = (\hat{T}+\hat{V}_{\rm ev}) + (\hat{V} - \hat{V}_{\mathrm{ev}}) \equiv \hat{H}_{\mathrm{ev}}+\delta\hat{V},
\ee
where we assume that $\hat{H}_{\mathrm{ev}}$ represents a nonperturbative problem that can 
be solved by means of QMC without the sign problem. By construction we 
assume that $\delta V$ can be regarded as a small correction 
to the energy that can be estimated in perturbation theory. To leading order we find
\be
E \lesssim \avg{\hat{H}} = \avg{\hat{H}_{\mathrm{ev}}}+\avg{\delta\hat{V}},
\label{eq:ELO}
\ee
and $\psi(\tau\to\infty)\sim\exp(-\tau \hat{H}_{\mathrm{ev}})\psi_{0}$ is the normalized ground state 
wavefunction of the evolution Hamiltonian. It is clear that our approach provides an upper bound for 
the ground state energy $E$ of the chiral Hamiltonian.

To construct the evolution Hamiltonian we note that each interaction that is spin-independent 
and attractive in momentum space ($V_{\rm ev}(q)\leqslant 0$) leads to a QMC simulation free 
from the sign problem (see for example~\cite{Koonin97}). Inspired by the one-boson exchange 
model, we express the evolution potential as (including the pion):
\be
%V_{\rm ev}(q) = \left( \frac{V_\pi}{m_\pi^2 + q^2} + \frac{V_\sigma}{m_\sigma^2 + q^2} + 
%\frac{V_\omega}{m_\omega^2+ q^2}\right) f(q),
V_{\rm ev}(q) = \sum_{\alpha=\pi,\sigma,\omega}\frac{V_\alpha}{m_\alpha^2c^2 + q^2} f(q),\label{evpot}
\ee
and  we apply a regulator function of the form $f(q)=\exp[-(q/\Lambda)^{30}]$.
These coupling constants and masses are fit (under the constraint that the sum is not positive) to 
minimize the expression 
\bea
\chi^2&=&\sum_{i,j} w^{(j)}
\left[\delta^{(j)}_{\rm EFT}(E_i)-\delta^{(j)}_{\rm ev}(E_i)\right]^2\nonumber\\
&&+\alpha\left[E_{\rm EFT}^{\rm (pert.)}-E_{\rm ev}^{\rm (pert.)}\right]^2
\label{chi2}
\eea
where $\delta^{(j)}(E_i)$ are phase shifts for partial waves
$j={}^1S_0,{}^3P_0,{}^3P_1,{}^3P_2$ obtained both for the chiral ${\rm N3LO}$ and 
the evolution potentials at given energy $E_i$, and the weights $w^{(j)}$ are respectively 
$1,\frac{1}{9},\frac{3}{9},\frac{5}{9}$. Thus the $P$-waves are weighted according their 
degeneracy. The range of phase shifts included in the fitting procedure is density dependent
and contains lab energies from $0$ to $\min[6E_{\rm lab}(k_F),350]$\,MeV 
where $E_{\rm lab}(k_F)=2k_F^2/M$ and $k_F$ is the Fermi momentum. A somewhat 
similar approach to handle the fermion sign problem was advocated in Ref.~\cite{Lynn2012}.

The role of the last term in Eq.\ (\ref{chi2})  is to ensure that the total energy of the 
neutron system interacting with the evolution potential and computed from
second-order perturbation theory $E_{\rm ev}^{\rm (pert.)}$ is the same as the energy 
computed with the chiral potential $E_{\rm EFT}^{\rm (pert.)}$ in the same framework. 
The ``stiffness'' of this requirement is governed by the parameter $\alpha$, and in practice we 
choose it so that differences in the perturbative estimates of the energy differ by less that 
$1\%$. We note also that this kind of evolution Hamiltonian is not necessarily unique, as
described in the {\it Results} section below.

Note that by construction the emerging evolution potential is density dependent, 
as in the case of the in-medium similarity renormalization group approach~\cite{Bogner2011}. 
Intuitively, it can be treated as an effective, in-medium
two-body interaction, where the density dependence accounts for
repulsive effects of nuclear three-body forces, Pauli blocking in the
medium, etc. In the zero-density limit the second term of Eq.~(\ref{chi2}) is negligible and 
the potential is fitted to phase shifts only and it reproduces correctly 
low energy scattering parameters, like scattering length or effective range. 

%-------------------------------------------------------------
\begin{figure}
\includegraphics[width=\pictsize\columnwidth]{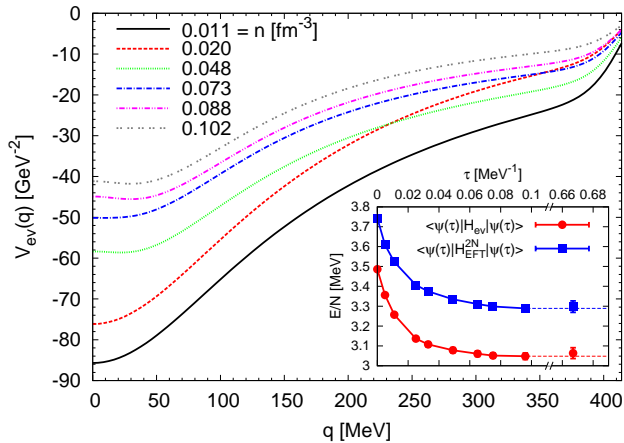}
\caption{ (Color online) Momentum-space evolution potentials (see Eq.\ (\ref{evpot})) 
employed in the imaginary-time propagation of the trial wavefunction, corresponding to different densities.
In the inset is shown the expectation values of the evolution potential (red solid circles) and 
the two-body chiral potential (blue squares) computed 
for the density $n=0.011\,{\rm fm^{-3}}$.
\label{fig:V_ev}}
\end{figure}
%-------------------------------------------------------------
Once the evolution potential is constructed, we generate the corresponding many-body 
wavefunction by means of AFQMC simulations. We 
consider a set of $N$ neutrons interacting on a three-dimensional cubic spatial lattice of 
extent $L=N_x^{} l$ and impose periodic boundary conditions. The lattice spacing 
$l=1.5\,{\rm fm}$ provides a natural ultraviolet cutoff scale, which we impose to be spherical 
in momentum space and consistent with the cutoff scale of the chiral theory, i.e.\ 
$\Lambda=p_{\rm cut}=\pi\hbar/l=414\,{\rm MeV/c}$. 
The imaginary-time evolution operator $\exp[-\tau\hat{H}_{\rm ev}]$ is expanded using 
a Trotter-Suzuki decomposition with temporal lattice spacing $\Delta \tau$, and the interaction 
$V_{\rm ev}$ is represented by means of a continuous Hubbard-Stratonovich (HS) transformation. 
In order to get faster convergence in the Monte-Carlo evaluation, we approximate the Gaussian 
quadrature emerging from the HS decomposition by a 5-points quadrature formula, which introduces 
an error that is small compared to that originating from the Suzuki-Trotter formula. 
The statistical error for Monte-Carlo quadrature estimation is below 1\%.

In this paper we work with lattice 
size $N_x=10$, which in previous studies of the unitary Fermi gas \cite{Bulgac08,Gabriel2013} led to 
systematic errors on the order of at most $\sim 10\%$. 
The main contribution to this error came from high momenta states beyond the momentum 
cut-off due to the slow decay of the universal high momentum tail in the occupation probability 
$n(p)\sim p^{-4}$. In the present work with chiral nuclear forces, the momentum distribution 
exhibits an exponential falloff (see {\it Results} section below), and therefore we expect 
improved finite-volume systematic errors. 
We have developed a new parallel code for these analyses and checked that calculations 
performed at zero temperature reproduce with sufficient accuracy the zero-temperature 
Bertsch parameter of the unitary Fermi gas. In particular, the superfluid gap of the unitary 
Fermi gas and the related properties are accurately reproduced. 
We consider densities from  $0.01\,{\rm fm^{-3}}$ to $0.10\,{\rm fm^{-3}}$, corresponding 
to particle numbers ranging from $38$ to $342$, thus larger than any previous calculations of 
neutron matter. In order to reduce the discretization errors, we work only with particle 
numbers corresponding to closed shells in the free Fermi gas model on the lattice.
Moreover, we have demonstrated the feasibility of exploring the 
nonperturbative properties of dilute neutron matter. We performed simulations with 38 particles in 
$12^3$, $14^3$ and $16^3$ boxes while keeping the lattice spacing fixed at $1.5\,{\rm fm^{-3}}$, which 
correspond to densities $0.0065$, $0.0041$ and $0.0028,{\rm fm^{-3}}$ respectively.

In addition to discretization errors and statistical errors (below $1\%$), our approach 
introduces another source of error, related to the fact that we approximate the ground-state 
wavefunction of the chiral Hamiltonian by the ground-state of the evolution Hamiltonian. The 
best strategy to quantify this error is to calculate the second-order correction in Eq.~(\ref{eq:ELO}). 
In this paper we show only that the first-order correction is small (at most $10\%$) and 
comparable to discretization errors. Assuming the perturbativeness of the expansion in Eq.~(\ref{eq:ELO}) 
we conclude that discretization errors are dominant in our approach, however, strict quantification 
will be subject of future studies. 

%%%%%%%%%%%%%%%%%%%%%%%%%%%%%%%%%%%%%%%%%%%%%%%
%%%%%%%%%%%%%%%%%%%%%%%%%%%%%%%%%%%%%%%%%%%%%%%
% Results
%%%%%%%%%%%%%%%%%%%%%%%%%%%%%%%%%%%%%%%%%%%%%%%
%%%%%%%%%%%%%%%%%%%%%%%%%%%%%%%%%%%%%%%%%%%%%%%
\ssec{Results}
In Fig.\ \ref{fig:V_ev} we plot the evolution potentials as a function of the momentum transfer $q$
for different densities obtained by minimizing the $\chi^2$ function in Eq.~(\ref{chi2}). Different 
initial choices for the coupling strengths and masses of the ``$\sigma$'' and ``$\omega$'' mesons 
resulted in nearly identical evolution potentials, except at the largest densities where variations
in the starting values gave a 2\% spread in the final energy per particle. We observe that the 
imposed energy constraint leads to a decrease in the overall 
strength of the evolution potentials as the density is increased. Physically this accounts for the 
presence of repulsive two- and three-body forces that become more important as the density
increases, so on average the total strength of the attractive nuclear potential must be reduced.

% %-------------------------------------------------------------
% \begin{figure}
% \includegraphics[width=\pictsize\columnwidth]{project.eps}
% \caption{ (Color online) Expectation values of the evolution potential (red solid circles) and 
% the two-body chiral potential (blue squares) computed over the normalized trial wave-function 
% $\psi(\tau)=\exp(-\tau \hat{H}_{\mathrm{ev}})\psi_{0}$ for density $n=0.011\,{\rm fm^{-3}}$.
% \label{fig:project}}
% \end{figure}
% %-------------------------------------------------------------
As an initial trial wavefunction we consider the Slater determinant of the lowest $N$ occupied 
discrete plane wave orbitals. The expectation values of the evolution Hamiltonian and the chiral
nuclear potential at imaginary time $\tau = 0$ are then simply the lattice Hartree-Fock energies. 
Deviations between the continuum Hartree-Fock predictions and those of the lattice were found 
to be at most a few percent when the particle number
corresponds to closed shells in the free Fermi gas model on the lattice.
In the inset of Fig.\ \ref{fig:V_ev}
we show the evolution in imaginary time of $\langle \psi(\tau) | \hat{H}_{\rm ev} | \psi(\tau) \rangle$ 
and $\langle \psi(\tau) | \hat{H}^{\rm 2N}_{\rm EFT} | \psi(\tau) \rangle$ for density $n=0.011\,{\rm 
fm^{-3}}$. Note that the left- and right-hand wavefunctions are evolved separately.
Typically we observe a very good convergence for imaginary times 
about $\tau\approx 0.1\,\rm{MeV}^{-1}$, which requires about 300 imaginary time steps.
Apart from a nearly constant shift, the imaginary-time 
dependence for both expectation values is very similar, indicating that our fitting procedure 
indeed produces the evolution potential, which correctly captures global features of the chiral potential. 

%-------------------------------------------------------------
\begin{figure}[t]
\includegraphics[width=\pictsize\columnwidth]{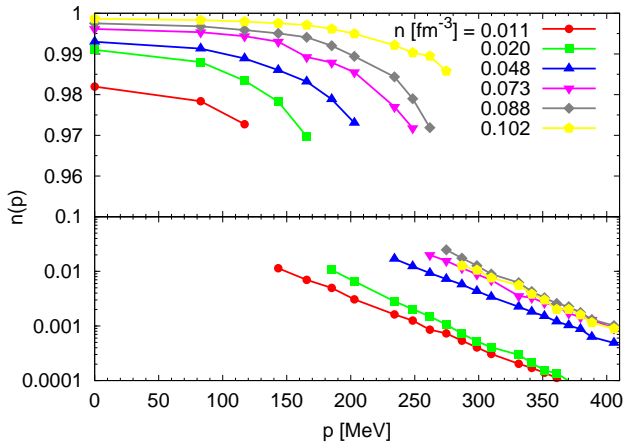}
\caption{ (Color online) Occupation probabilities of neutron matter as a function of momentum 
for selected densities.}
\label{fig:occup2}
\end{figure}
%-------------------------------------------------------------
Our calculation procedure gives us access to the wave function 
in both the coordinate and momentum representation. In Fig.\ \ref{fig:occup2} we show the 
momentum distribution associated with the evolution Hamiltonian $\hat{H}_{\rm ev}$ for pure neutron 
matter at selected densities. As the density increases and the evolution Hamiltonian weakens, 
the depletion in the occupation probability at low momenta is reduced. In all simulations the 
single-particle occupation probabilities for the highest energy states is below one percent.

%-------------------------------------------------------------
\begin{figure}[t]
\includegraphics[width=\pictsize\columnwidth]{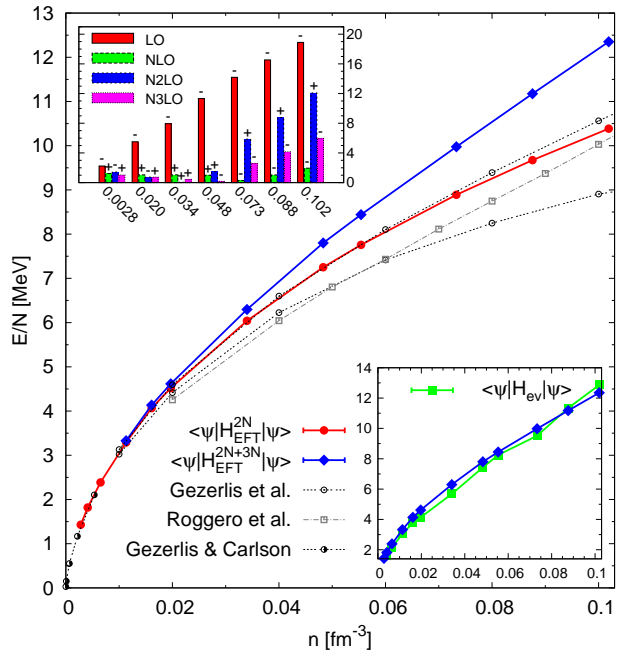}
\caption{(Color online) Equation of state of pure neutron matter calculated using AFQMC
with the N3LO chiral two-nucleon potential (red circles) plus the N2LO three-nucleon 
contribution (blue diamonds). For comparison we show the results of Gezerlis \textit{et al.}~\cite{Gezerlis13}, 
Roggero \textit{et al.}~\cite{Roggero14} and Gezerlis \& Carlson~\cite{GezerlisCarlson} of QMC 
calculations with two-body forces alone.
In the upper inset we show the contributions to the energy per particle from different orders in the chiral
expansion (``+'' and ``$-$'' refer to repulsive and attractive components, respectively). 
The lower inset demonstrates that the last term in Eq.~(\ref{eq:ELO}) is perturbative.
% Lower-right inset: comparison between the expectation values of $\hat{H}_{\rm ev}$ and 
% $\hat{H}_{\rm EFT}$ in the correlated many-body ground state. 
\label{fig:E_vs_n}}
\end{figure}
%-------------------------------------------------------------

In Fig.\ \ref{fig:E_vs_n} we present AFQMC results for the equation of
state of pure neutron matter~\cite{EPAPS}. Evaluating only the chiral two-nucleon
force in the correlated ground state (shown in red solid circles), we
find that the equation of state is consistent with previous quantum
Monte Carlo simulations employing N2LO chiral 2N interactions
\cite{Gezerlis13,Roggero14}. At very low densities our results
match perfectly onto the QMC results obtained with the effective interaction which captures
correctly only scattering length and effective range~\cite{GezerlisCarlson}. 
At these low densities the pairing correlations are very strong 
($\Delta/\varepsilon_F \approx 1/4$, where $\varepsilon_F$ is the Fermi energy
of a free neutron gas with the same density)
and the nice agreement with~\cite{GezerlisCarlson} demonstrates that we capture them very accurately.
The chiral nuclear potential most similar
the one employed in the current study is the local 400 MeV cutoff
potential of Ref.\ \cite{Gezerlis13} (see the lower line of gray open circles in Fig. \ref{fig:E_vs_n}). 
Our results for the energy per particle
are more repulsive by about 1 MeV, and although further investigations
are required, we note that the chiral nuclear interaction in Ref.\
\cite{Gezerlis13} is more strongly attractive in relative $P$-waves
than the potential we employ. Computing also the expectation value of
the N2LO three-nucleon force over the evolved wavefunction introduces
significant additional repulsion above $n=0.02$\,fm$^{-3}$, as seen
from the solid blue diamonds in Fig.\ \ref{fig:E_vs_n}. Differences
between the expectation value of the evolution Hamiltonian and the
full chiral nuclear 2N + 3N interaction (which can be regarded as the
first-order correction to the energy in perturbation theory) are small
as shown in the lower-right inset to Fig.\ \ref{fig:E_vs_n}. In the
upper-left inset, we show the expectation value of the chiral
Hamiltonian decomposed according to the chiral order.
%The attractive 
%components are labeled with ``$-$'' while the repulsive components are labeled with ``+''.

In the above calculations we translate the lattice results to the continuum limit with the 
following procedure: i)~from the lattice simulations we extract the dimensionless quantity 
$\frac{\avg{\hat{O}}}{\avgo{\hat{O}}}$, where $\psi$ is the ground state of the evolution 
Hamiltonian, $\psi_0$ is the free Fermi gas wave function, and both expectation values are 
computed on the lattice, ii)~to convert the lattice result into a dimensionful quantity we multiply 
by $\langle \psi_0 | \hat O | \psi_0 \rangle^{(\rm cont.)}$, computed
in the continuum limit. %Finite-size errors are estimated 

%%%%%%%%%%%%%%%%%%%%%%%%%%%%%%%%%%%%%%%%%%%%%
%%%%%%%%%%%%%%%%%%%%%%%%%%%%%%%%%%%%%%%%%%%%%
% Conclusions
%%%%%%%%%%%%%%%%%%%%%%%%%%%%%%%%%%%%%%%%%%%%%
%%%%%%%%%%%%%%%%%%%%%%%%%%%%%%%%%%%%%%%%%%%%%

% \ssec{Conclusions}
% We have presented calculations of the cold neutron matter equation of
% state based on chiral two- and three-nucleon forces. A novel
% auxiliary-field quantum Monte Carlo method capable of simulating
% large numbers of nucleons on the lattice has been introduced that
% employs evolution Hamiltonians free of the fermion sign
% problem. The method is well suited for the inclusion of N3LO three-
% and four-body forces and for future studies of finite nuclei.

%%%%%%%%%%%%%%%%%%%%%%%%%%%%%%%%%%%%%%%%%%%%%
%%%%%%%%%%%%%%%%%%%%%%%%%%%%%%%%%%%%%%%%%%%%%
% Acknowledgments
%%%%%%%%%%%%%%%%%%%%%%%%%%%%%%%%%%%%%%%%%%%%%
%%%%%%%%%%%%%%%%%%%%%%%%%%%%%%%%%%%%%%%%%%%%%
\ssec{Acknowledgments}
We are grateful to P.\ Magierski and R.\ Machleidt for helpful discussion, and we 
thank A.\ Gezerlis and A.\ Roggero for sharing the results of their numerical simulations.
This work was supported in part by US DOE Grant No.\ DE-FG02-97ER-41014 and the Polish National Science Center (NCN)
under Contracts No. UMO-2013/08/A/ST3/00708 and
No. UMO-2012/07/B/ST2/03907.  
Calculations reported here have been performed at the University of
Washington Hyak cluster funded by the NSF MRI Grant No.\@ PHY-0922770 
and at the Interdisciplinary Centre for Mathematical and Computational 
Modelling (ICM) at University of Warsaw. G.W.\@ acknowledges the Center 
for Advanced Studies at Warsaw University of Technology for
the support under Contract No. 58/2013 (international research 
scholarships financed by the European Union from the European Social 
Funds, CAS/32/POKL). KJR was supported by the DOE Office of Science, Advanced 
Scientific Computing Research, under award number 58202 ``Software Effectiveness 
Metrics" (Lucille T. Nowell). 

%%%%%%%%%%%%%%%%%%%%%%%%%%%%%%%%%%%%%%%%%%%%%
%%%%%%%%%%%%%%%%%%%%%%%%%%%%%%%%%%%%%%%%%%%%%
 
% \bibliography{biblio}{}

\end{document}